\documentclass[onecolumn]{aastex631}

\usepackage{natbib}
\usepackage{bibentry}
\usepackage{graphicx}
\usepackage{epsfig}
\usepackage{epstopdf}
\usepackage{subfigure}

\newcommand{\kiauhoku}{\texttt{kiauhoku}}

\begin{document}

\title{A Lack of Mass-Gap Compact Object Binaries in APOGEE}

\author[0009-0007-9766-2324]{Meir Schochet}
\affiliation{Department of Astronomy, University of Florida, Bryant Space Science Center, Stadium Road, Gainesville, FL 32611, USA}

\author[0000-0002-4818-7885]{Jamie Tayar}
\affiliation{Department of Astronomy, University of Florida, Bryant Space Science Center, Stadium Road, Gainesville, FL 32611, USA}

\author[0000-0001-5261-3923]{Jeff J. Andrews}
\affiliation{Department of Physics, University of Florida, 2001 Museum Road, Gainesville, FL 32611, USA}
\affiliation{Institute for Fundamental Theory, University of Florida, 2001 Museum Road, Gainesville, FL 32611, USA}

\date{June 12, 2024}

\begin{abstract}
Depending principally on mass, the compact object remnant left behind after a star's life may be a white dwarf (WD), neutron star (NS), or black hole (BH). While we have large samples of each of these remnants, we lack knowledge of the exact conditions separating these outcomes. The boundary between low-mass BHs and massive NSs is particularly poorly understood, as few objects between 2-5 $M_\odot$ are known. To probe this regime, we search the APOGEE DR17 dataset of 657,000 unique stars for binary systems with one stellar component that exhibit large radial velocity shifts over multiple observations. We identify 4751 likely binary systems, and estimate a minimum mass for each system's ``invisible companion" under the assumption of tidal synchronization. Two systems have companion masses $\gtrsim$ 2 $M_\odot$, although we conclude that neither are good candidates for possessing a mass-gap NS or BH companions.
\end{abstract}

\section{Introduction}
Our knowledge of how massive stars end their lives is incomplete. One reason is that isolated NSs and BHs are faint and difficult to observe, especially compact objects in the ``mass-gap" between 2-5 $M_\odot$ \citep{Fryer2012}. Binary population synthesis models predict that objects in the regime form from zero-age main sequence progenitors between 20 to 30 $M_\odot$ \citep{Belczynski_most_2008}, although there are large uncertainties including the metallicity dependence of stellar winds \citep{Mapelli2013} and the maximum mass of NSs \citep{Siegal_2023}. There have been extensive efforts to bolster the sample of known NSs and BHs, and through binary interactions we can observe them \citep{Thompson2019}. Our approach is to search for mass-gap compact objects by looking at single lined spectroscopic binaries with large radial velocity (RV) variations.

\section{Methods}
\subsection{APOGEE Data Reduction Pipeline}
For our analysis we use the 17th Data Release \citep[DR17;][]{Abdurrouf2022} of the Sloan Digital Sky Survey's \citep[SDSS;][]{Majewski2017} Apache Point Observatory Galactic Evolution Experiment (APOGEE). APOGEE collects near-infrared spectra and calculates stellar parameters and abundances from multiple observations. DR17 objects were reduced using the Doppler \citep{Nidever2015}, which fits more accurate RVs, stellar parameters, and abundances compared to previous data releases. Doppler also flags the number of stellar components it identifies while fitting, and we select 503,451 objects with multiple visits and one spectral component. We further select 89,798 systems with temperatures between 4100 K $<$ $T_{\rm eff}$ $<$ 7000 K and log(g) between 3.6 dex $<$ $\log$(g) $<$ 4.5 dex. This ensures precise stellar parameters and rotational velocities, while also removing subsolar-mass dwarf stars that may host low-mass companions. We pick 8837 stars with APOGEE v$_{\rm scatter}$ $\geq$ $1$ km s$^{-1}$ as an indicator of binarity, and finally select 4751 systems with $v \sin(i)>$ 10 km s$^{-1}$ as evidence of tidal spin-up from a close companion \citep{Tayar2015}.

\subsection{Companion Mass Estimates}
We estimate a minimum mass for invisible companions to each star in our sample using:
\begin{equation}
f = \frac{M^3_{2} \sin^3(i)}{(M_{2} + M_{1})^2} = \frac{P_{orb} K^3}{2 \pi G}(1-e^2)^\frac{3}{2}.
\end{equation}
Here, $M_{2}$ is the unseen companion's mass and $M_{1}$ is the stellar component's mass. We use \kiauhoku{} \citep{Claytor2020} to infer our $M_{1}$ masses from MIST models \citep{Choi2016} using APOGEE $T_{\rm eff}$, $\log$(g), and [M/H] values, assigning a mass of 1 $M_\odot$ to any stars with no model mass returned. We assume an edge-on configuration and solve using two methods. In the first, we assume tidal synchronization and a circular orbit, we use v$_{\rm scatter}$ as a proxy for velocity semi-amplitude, and estimate a period using APOGEE $v \sin$(i) and radius values from Gaia and SED fits \citep{Yu2023}. This tidal synchronization procedure returns 4288 estimates. The second uses Keplerian orbit parameters from the Joker Value Added Catalog \citep{PriceWhelan2017}. The Joker takes RVs and generates convergent posterior samplings for orbital parameters, allowing us to estimate masses for a separate 1798 objects. We note that most of our sample has $\leq$ 8 visits which may not be sufficient for the Joker to produce reliable fits. We nevertheless include these estimates in our catalog in Zenodo at \dataset[doi: 10.5281/zenodo.10901389]{https://doi.org/10.5281/zenodo.10901389} for completeness.

\section{Analysis}
\subsection{Sample Validation and Analysis}
Figure (1) panel (a) shows a previously identified mass-gap object, alongside our most interesting candidate in panel (b). The regime in which our candidates are found is shown in panel (c), and panel (d) demonstrates that our estimates are more alike for the two procedures when stars have more visits. Panel (e) shows the final distribution of our tidal synchronization mass estimates, confirming a lack of companions between 2-5 $M_\odot$, except for our best candidate (Sec. 3.2). Finally, panel (f) shows the binary fraction across metallicity for our 4751 likely binaries compared to the 89,798 well-measured systems within our regime of $T_{\rm eff}$ and $\log$(g), showing an anti-correlation of metallicity to binary fraction similar to \citet{Moe2019}.

\subsection{Best Candidate}
Our most promising system is 2MASS J19245871$+$4444081. This candidate has five LAMOST RVs which demonstrate $\Delta$RV$_{\rm LAMOST, max} \approx$ 66 km s$^{-1}$ from 2013-2020, and four APOGEE RVs showing $\Delta$RV$_{\rm APOGEE, max} \approx$ 2$\Delta$RV$_{\rm LAMOST, max}$ over three days in 2016. This candidate has a synchronization estimated companion mass of 2.693 $M_{\odot} \sin(i)$ and a Joker estimated mass of 2.367 $M_{\odot} \sin(i)$. Upon further investigation, the radius from \citet{Yu2023} used for the tidal synchronization estimate (R $\approx$ 15 $R_{\odot}$) was substantially too large for the masses interpolated from \kiauhoku{} grids. Instead, we adopt radius (2.65 $R_{\odot}$) and $M_{1}$ values (1.29 $M_{\odot}$) from MIST grids for a re-estimated tidal synchronization mass of 1.136 $M_{\odot} \sin(i)$ and a new $\frac{P}{\rm \sin(i)}$ of $\approx$ 4.4 days. The candidate has TESS and Kepler lightcurves which constrain the period to $\approx$ 1.8 days and give a final synchronization estimate of 0.727 $M_{\odot} \sin(i)$.  The candidate therefore may host a WD companion; however no GALEX, Chandra, eROSITA, or XMM data was available to confirm any high energy excess.

\section{Conclusions}
Consistent with other works, we find that mass-gap compact objects are rare. We find that in general, assuming tidal synchronization produces a catalog of estimated masses that follow the expected trends in mass and metallicity. We identify one interesting candidate, and are optimistic that future large surveys like the MWM \citep{Kollmeier} will reliably identify many companions, perhaps including rare mass-gap compact objects.

\section{Acknowledgements}
M.S. thanks the University Scholars Program at UF. We utilize data from SDSS (\href{https://www.sdss.org/collaboration/citing-sdss/}{SDSS Website}).

\pagebreak
\bibliographystyle{aasjournal} 

{}

\pagebreak
\begin{figure}
\gridline{\leftfig{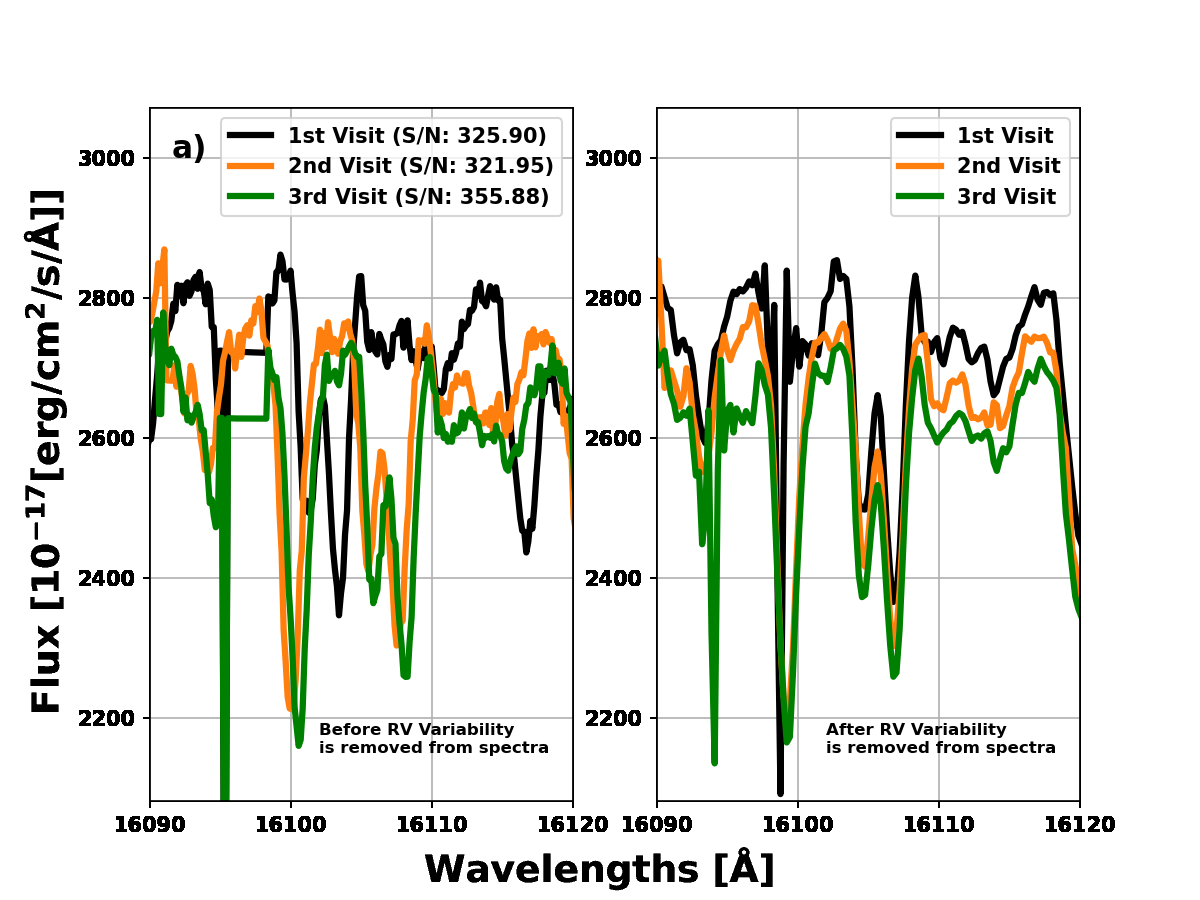}{0.49\textwidth}{} \rightfig{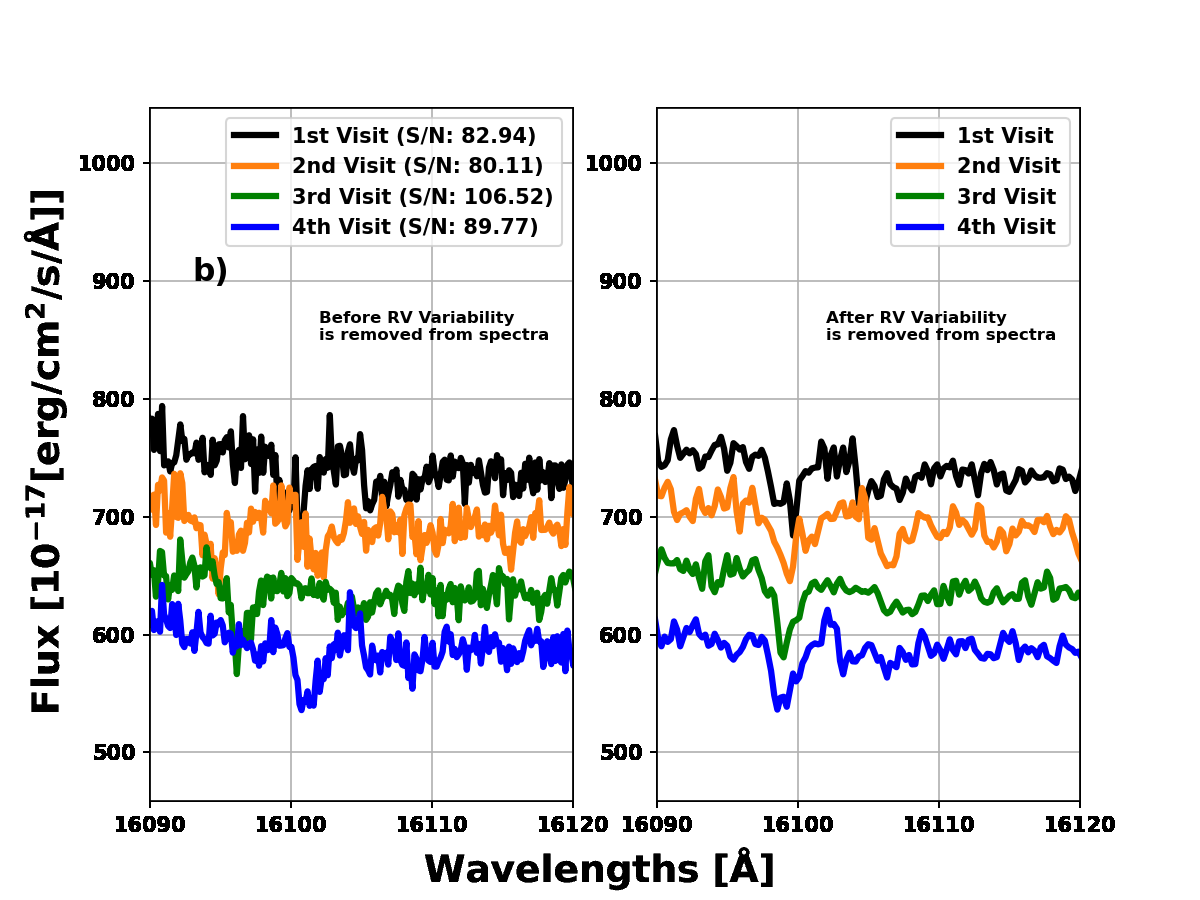}{0.49\textwidth}{}}
\gridline{\leftfig{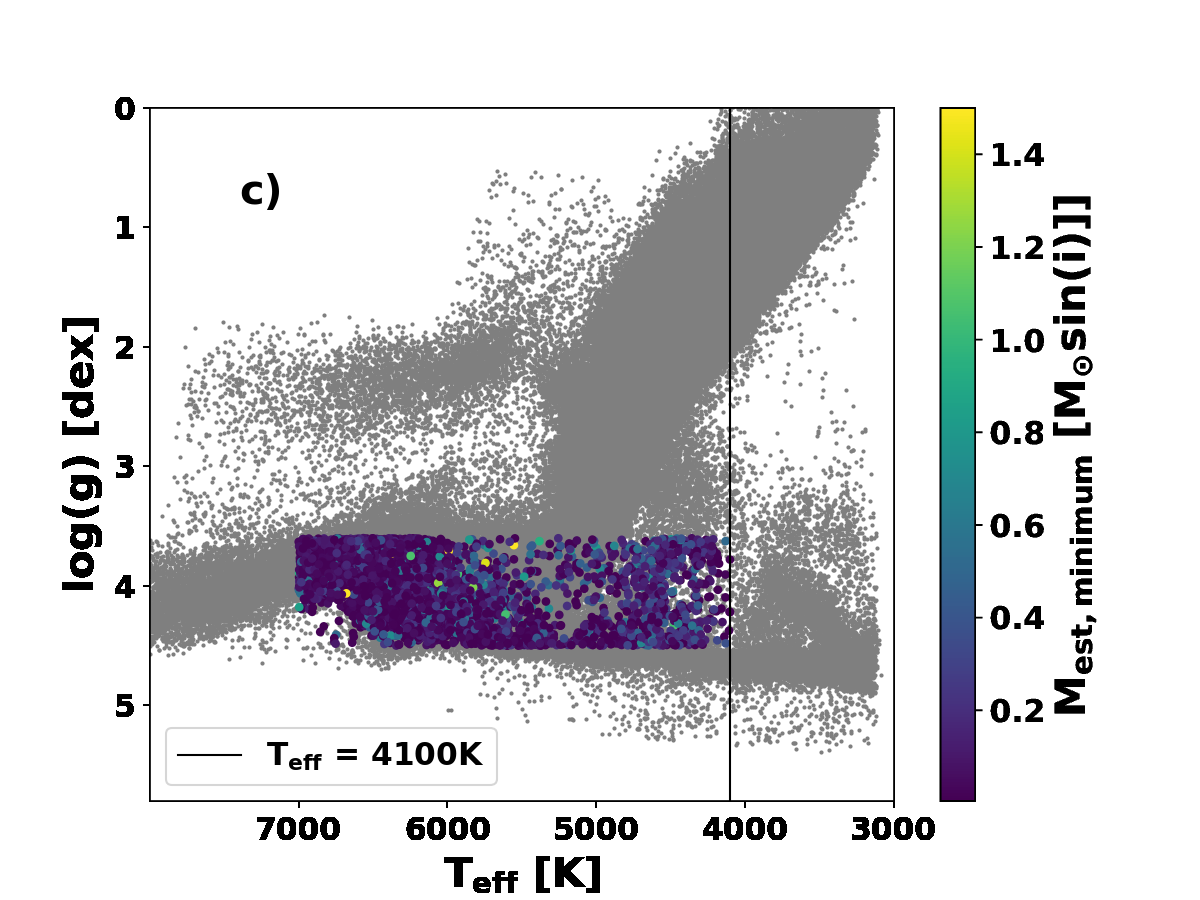}{0.49\textwidth}{} \rightfig{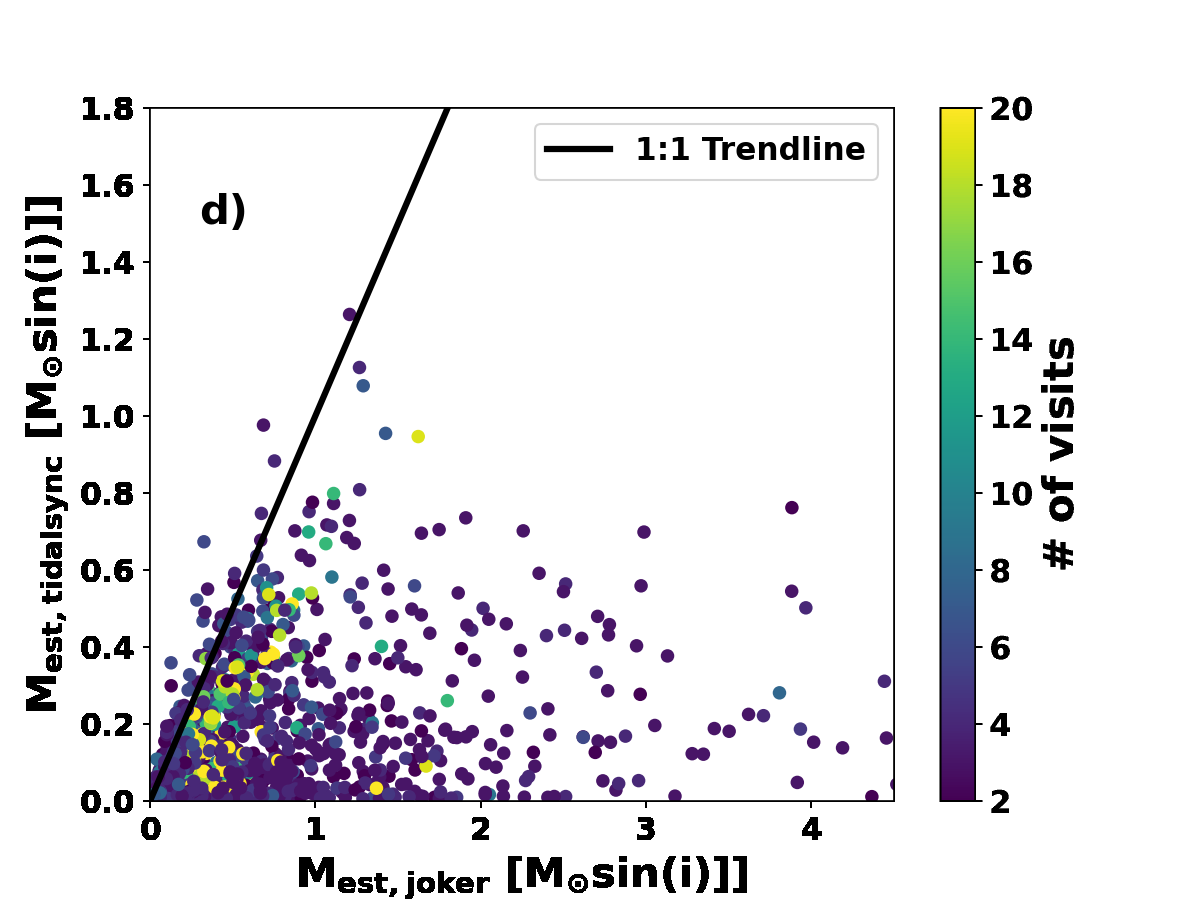}{0.49\textwidth}{}}
\gridline{\leftfig{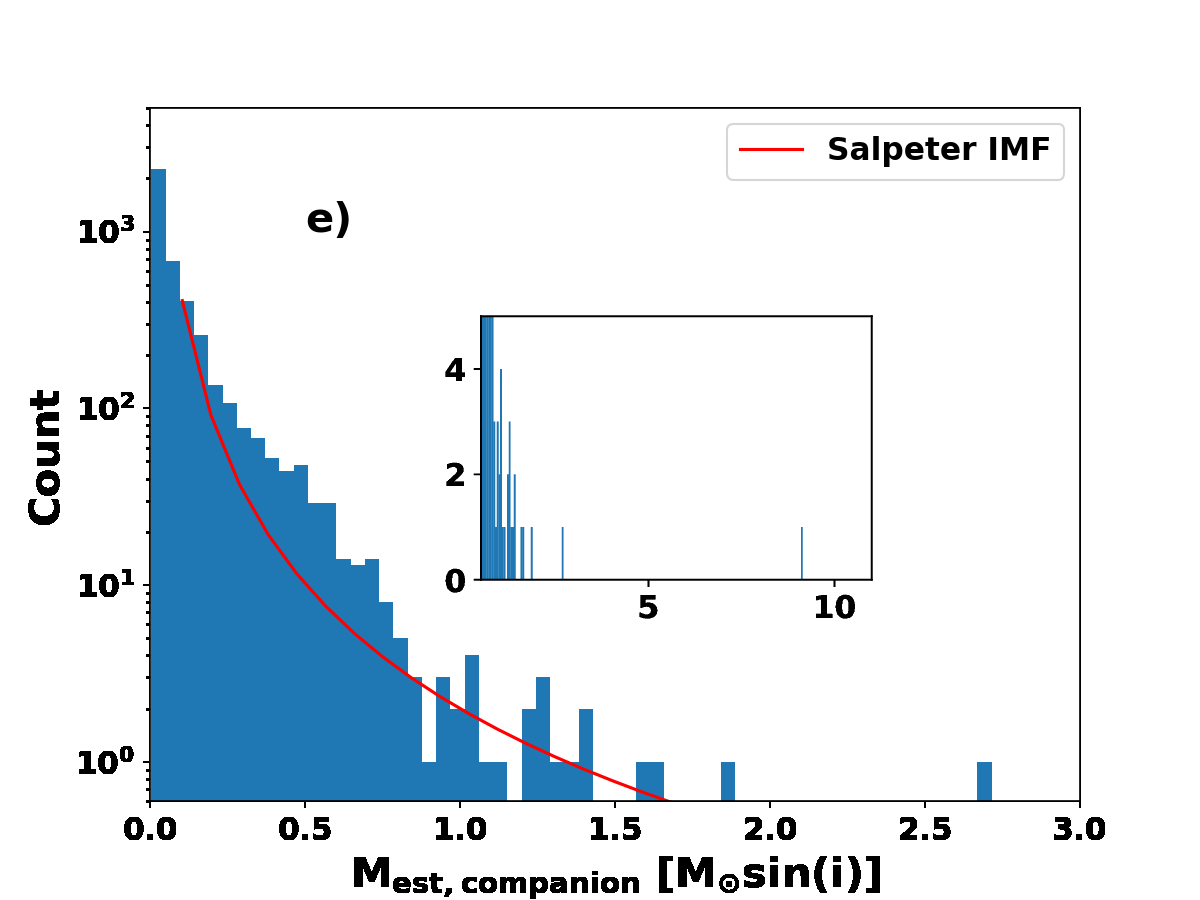}{0.49\textwidth}{} \rightfig{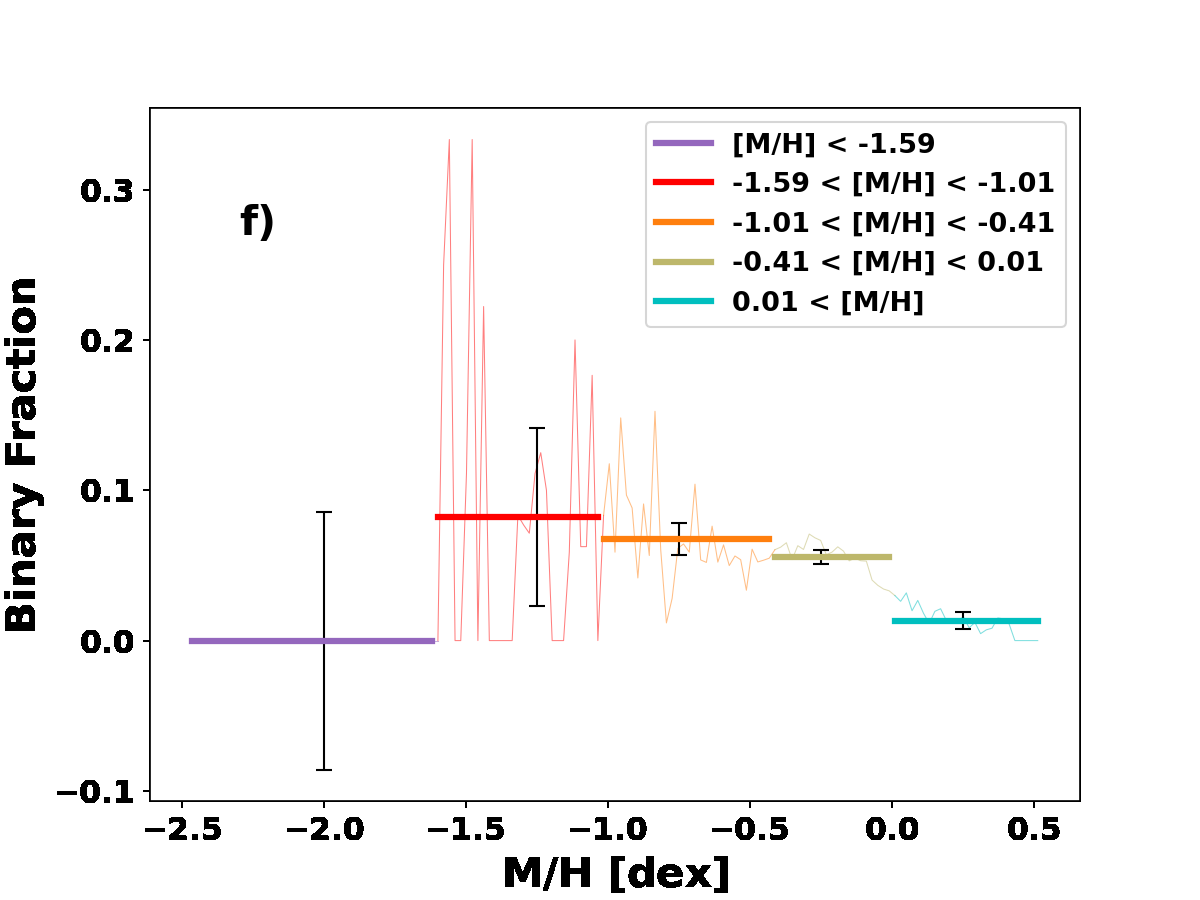}{0.49\textwidth}{}}
\caption{We show that our pipeline can identify RV variability from Doppler shifts in previously (a, \citeauthor{Thompson2019}) and newly (b) identified systems. We show the regime (c) where our systems are found and demonstrate that our different mass estimates do not completely agree (d). We show that our detected companions are consistent with expected-trends with mass (e) and metallicity (f).}
\end{figure}
\end{document}